\documentstyle[12pt,russian]{article}
\def\bea{\begin{eqnarray}}
\def\eea{\end{eqnarray}}

\def\beq{\begin{equation}}
\def\eeq{\end{equation}}
\def\ba{\beq\new\begin{array}{c}}
\def\ea{\end{array}\eeq}
\def\be{\ba}
\def\ee{\ea}

\makeatletter
\newdimen\normalarrayskip % skip between lines
\newdimen\minarrayskip % minimal skip between lines
\normalarrayskip\baselineskip
\minarrayskip\jot
\newif\ifold \oldtrue \def\new{\oldfalse}
\def\arraymode{\ifold\relax\else\displaystyle\fi} % mode of array entries
\def\eqnumphantom{\phantom{(\theequation)}} % right phantom in eqnarray
\def\@arrayskip{\ifold\baselineskip\z@\lineskip\z@
\else
\baselineskip\minarrayskip\lineskip2\minarrayskip\fi}
\def\@arrayclassz{\ifcase \@lastchclass \@acolampacol \or
\@ampacol \or \or \or \@addamp \or
\@acolampacol \or \@firstampfalse \@acol \fi
\edef\@preamble{\@preamble
\ifcase \@chnum
\hfil$\relax\arraymode\@sharp$\hfil
\or $\relax\arraymode\@sharp$\hfil
\or \hfil$\relax\arraymode\@sharp$\fi}}
\def\@array[#1]#2{\setbox\@arstrutbox=\hbox{\vrule
height\arraystretch \ht\strutbox
depth\arraystretch \dp\strutbox
width\z@}\@mkpream{#2}\edef\@preamble{\halign
\noexpand\@halignto
\bgroup \tabskip\z@ \@arstrut \@preamble \tabskip\z@ \cr}%
\let\@startpbox\@@startpbox \let\@endpbox\@@endpbox
\if #1t\vtop \else \if#1b\vbox \else \vcenter \fi\fi
\bgroup \let\par\relax
\let\@sharp##\let\protect\relax
\@arrayskip\@preamble}
\def\eqnarray{\stepcounter{equation}%
\let\@currentlabel=\theequation
\global\@eqnswtrue
\global\@eqcnt\z@
\tabskip\@centering
\let\\=\@eqncr
$$%
\halign to \displaywidth\bgroup
\eqnumphantom\@eqnsel\hskip\@centering
$\displaystyle \tabskip\z@ {##}$%
\global\@eqcnt\@ne \hskip 2\arraycolsep
$\displaystyle\arraymode{##}$\hfil
\global\@eqcnt\tw@ \hskip 2\arraycolsep
$\displaystyle\tabskip\z@{##}$\hfil
\tabskip\@centering
&{##}\tabskip\z@\cr}
\begingroup\ifx\undefined\newsymbol \else\def\input#1 {\endgroup}\fi

\begin{document}

\setcounter{footnote}{1}
\def\thefootnote{\fnsymbol{footnote}}

\begin{center}
\hfill ITEP/TH - 81/02\\
%\hfill hep-th/\\
\vspace{0.3in}
{\Large\bf }
\end{center}

\begin{center}
{\large\bf INTEGRABLE SYSTEMS, OBTAINED BY POINT FUSION FROM
RATIONAL AND ELLIPTIC GAUDIN SYSTEMS}
%\rule{5cm}{1pt}
\end{center}

\bigskip

\centerline{{\large Chernyakov Yu. } \footnote{Institute of Theoretical and
Experimental Physics, Moscow,
Russia, e-mail: chernyakov@gate.itep.ru }}

\abstract{\footnotesize Using the procedure of the marked point
fusion, there are obtained integrable systems with poles in the
 matrix of the Lax operator order higher than one, considered
Hamiltonians, symplectic structure and symmetries of these
systems. Also, taking the Inozemtsev Limit procedure it was found
the Toda-like system having nontrivial commutative relations
between the phase space variables.}

%\bigskip
%\setcounter{footnote}{0}

\section{Introduction}
\paragraph{Rational and Elliptic Gaudin systems}
In the present work there were used Hamiltonians, Lax operators
and symplectic structure of the rational and elliptic Gaudin
systems. These systems were studied in the works $[1]-[4]$. Let us
give their short description.

Let us begin from rational Gaudin system and consider $n$ marked
points $x_{a}, a=\overline{1,n}$ on $CP^{1}$ and assign a
coadjoint orbit of the group $SL(N,C)$ to each point. The
coordinates $p_{a}^{ij}$ are functions on the corresponding
orbits. The Lie-Poisson brackets of $p_{a}^{ij}$ (spins) have the
following form (Appendix B.):
  \be \{ p_{a}^{ij},
p_{b}^{kl} \} = \delta_{ab} ( \delta^{il} p_{a}^{kj} - \delta^{kj}
p_{a}^{il} ) \ee On the phase space, which is the direct product
of the orbits factorized by $SL(N,C)$ group with coordinate
independent elements on $CP^{1}$, it is possible to define the
integrable system. The Hamiltonians of this system commuting with
respect to the Lie-Poisson brackets (1) have the following form:
\be
       H_{a} = \sum_{b \neq a} \frac{<p_{a}p_{b}>}{x_{b}-x_{a}},\\
\ee where $< >$ means the trace
. Other Hamiltonians are also expressible in terms of the traces
of the spin products and differences of coordinates at the marked
points in certain degrees.

Let us describe now the elliptic case: instead of $CP^{1}$ let us
consider the elliptic curve. The integrable system is determined
on the symplectic factor-space $R$ ($[4]$) which has the following
form: \be

R = (u,v)\times(\prod_{a} {\mathcal O}_{a}//H),

\ee where ${\mathcal O}_{a}$ - the
coadjoint orbit of the $SL(N,C)$ group representation,\\
 $H$ - Cartan subgroup,\\
$u,v \in \mathcal H$ - Cartan subalgebra.

The commutation relations between $v$,$u$ and $p_{s}^{ij}$ in
addition to those between $p_{a}^{ij}$ have the following form:
 \be \{ v^{i}, u^{j} \} = \delta^{ij},\ \ \

\{ v^{k}, p_{a}^{ij} \} = 0, \ \ \

\{ u^{k}, p_{a}^{ij} \} = 0 \ee Well-known Hamiltonian of the
elliptic Gaudin system describes a physical system consisting of
$n$ interacting particles: \be H = \sum_{i} \frac{1}{2}(v^{i})^{2}
 + \sum_{i \neq j} <p^{2}> \cdot E_{2}(u^{ij}),
\ee
where $v_{i}$ - momenta,\\
$u^{ij}=u^{i}-u^{j}$ - difference of particle coordinates,\\
$E_{2}(u^{ij}) \equiv E_{2}(u^{ij},\tau)$ - interaction potential
of particles, represented by the elliptic Eisenstien functions.
This function is connected with $\wp$-Weirstrass function:
 $E_{2}(z,\tau) = \wp(z,\tau) + 2 \zeta(\frac{1}{2})$, where $z$ -
 coordinate on the torus. Function $E_{2}(z,\tau)$ is expressed in
 terms of $\theta$-function ( Appendix L. ):
 $E_{2}(z,\tau) = - \partial^{2}_{z} \ln (\theta(z,\tau))$.

Note that in the previously mentioned works there are studied
systems which are the examples of the Hitchin systems - integrable
systems on the cotangent bundles to the moduli spaces of
holomorphic bundles over Riemann curves. These systems may be
obtained by the symplectic reduction ($[4],[5]$) from the input
phase spaces. Using the Lax operator we get the important way to
describe the integrable systems. So, in the case of rational
Gaudin system the Lax operator takes the following form: \be
 L(z)= \sum_{a} \frac{p_{a}}{(z-x_{a})},
\ee
     where $z$ - coordinate on $CP^{1}$.

 The matrix of the Lax operator in the elliptic case takes the
 following form:

diagonal part -
 \be L^{ii} = v^{i} + \sum_{a} p_{a}^{ii}
\cdot E_{1}(z-x_{a}) \ee

non-diagonal part - \be

 L^{ij} = \sum_{a} p_{a}^{ij} \cdot
 \exp  \left [- 2 \pi i u^{ij} \cdot
\frac{ (z - x_{a}) - (\bar{z} - \bar{x_{a}}) }{ \tau - \bar{\tau}
} \right ] \cdot \varphi (u^{ij}, z - x_{a}), \ee

where: \be \varphi (u^{ij}, z - x_{a}) =
 \frac{ \theta(u^{ij} + z - x_{a})\theta^{'}(0)}{ \theta(u^{ij}) \theta(z -
 x_{a})}. \ee

In the case of the Hitchin system the matrix of the Lax operator coincides
with the solution of the moment map equation appearing in the symplectic
reduction procedure. The Hamiltonians of the integrable system come from the
appropriate basis expansion. The Lax operators studied in the
above-mentioned works admit the first order poles.

In the present work we try to answer the naturally arising
questions: how do the systems with the higher poles in the Lax
operator look like, what kind of symplectic form do these systems
have and what are the moduli spaces of these systems.

\paragraph {Results and methods being used to obtain the systems}

We use the procedure of the point fusion to obtain the integrable
systems from rational and elliptic Gaudin systems. The main idea
of this method consists of finding such decomposition of the
variables $p$, which gives us the pole order in the matrix of Lax
operator being higher than one at some marked point, for example
$x_{a}$, after taking the limit $x_{b}-x_{a}=\varepsilon \rightarrow 0$,
 where $\varepsilon$ is the parameter of the point
fusion( we put here the coefficient at $\varepsilon$ equal to 1).
It implies the existence of Hamiltonian and Casimir numbers in
this new system being the same as in the initial one. This
situation is realized by passing to the new variables $p^{r}$
using grading decomposition: \be p_{a} = \alpha_{0} p^{0} +
\alpha_{1} p^{1} \varepsilon^{-1} + ... + \alpha_{r} p^{r}
\varepsilon^{-r},
 \ee
 where $\alpha_{r}$ - some variables.

The Hamiltonians in the case of two point (chosen from $n$ marked
point) fusion ($x_{b} \rightarrow x_{a}$) have the following form:
\be

H^{2,2}_{1} = <(p^{0})^{2}> + 2\sum_{c \neq a,b} \frac{<p^{1}
p_{c}>}{(x_{a} - x_{c})}, \ \

H^{2,1}_{1} = \sum_{c \neq a,b}^{n} \left[ \frac{<p^{0}
p_{c}>}{(x_{a} - x_{c})} - \frac{<p^{1}p_{c}>}{(x_{a} -
x_{c})^{2}} \right],

\\

H^{2,1}_{c} =  2\frac{<p^{1} p_{c}>}{(x_{c} - x_{a})^{2}} +
\frac{<p^{0} p_{c}>}{(x_{c} - x_{a})} + \sum_{d \neq c,a,b}
\frac{<p_{c} p_{d}>}{(x_{c} - x_{d})},

\ee

where $< >$ means the trace. The commutation relations in this case
are determined in chapter 2, formula (22). The Lax operators of
these systems were described in work $[6]$. Note, that in the case
of $r$ point (chosen from $n$ marked point) fusion the Lax
operator has the following form: \be

L_{new} = \frac{p^{r} \cdot k_{r} }{(z - x_{a})^{r}} + ... +
\frac{p^{1} \cdot k_{1} }{(z - x_{a})^{2}} + \frac{p^{0} \cdot
k_{0} }{(z - x_{a})} + \sum_{c \neq a,b} \frac{p_{c}}{(z-x_{c})},

\ee where $k_{r}$ - some coefficients depending on the parameters
of decomposition $p^{r}$, the parameters of the point fusion and
some conditions arising from the requirement of the absence of
singularities in the final expression.  There is a correspondence
between $p^{r}$ and upper triangular matrices which are isomorphic
to polynomial algebra of $\varepsilon^{-1}$ variable with
coefficients $p^{r}$. Commutation relations between $p^{r}$ were
found in chapter 2 (table (28)). Note, that special case of the Lax operator
(12) was obtained in the work $[7]$.

In the elliptic case the Hamiltonians of the two point fusion have
the following form($p \in sl(2,C)$):
 \be

H^{2,2} =

 - (p^{0})^{12}(p^{0})^{21} +
 E_{1}(u) \cdot
[ - (p^{0})^{12}(p^{1})^{21} + (p^{0})^{21}(p^{1})^{12} ]+\\

+ (p^{1})^{12}(p^{1})^{21} \cdot [ E_{1}^{'}(u) + E_{1}^{2}(u)]

- 2v(p^{1})^{11}\\

H^{2,0} = v^{2} + ((p^{1})^{11})^{2} \cdot [\frac{1}{6}
E_{1}^{'''}(u) + (E_{1}^{'}(u))^{2}]

+ (p^{0})^{12}(p^{0})^{21} \cdot E_{1}^{'}(u) +\\

+ [(p^{0})^{12}(p^{1})^{21} - (p^{0})^{21}(p^{1})^{12}] \cdot

[E_{1}(u)E_{1}^{'}(u) + \frac{1}{2}E_{1}^{''}(u)

] -\\

- (p^{1})^{12}(p^{1})^{21} \cdot

[(E_{1}^{'}(u))^{2} + E_{1}^{2}(u)E_{1}^{'}(u) +
E_{1}(u)E_{1}^{''}(u) + \frac{1}{3}E_{1}^{'''}(u)],

 \ee
where $u,v \in \mathcal H$ belong to Cartan subalgebra,\\
$E_{1}(u^{ij}) \equiv E_{1}(u^{ij},\tau)$
 are the elliptic Eisenstien functions ( Appendix A. ) determined
 on the complex torus $T^{2}$ with modulus $\tau$.
 This function is connected with $\zeta$-Weirstrass function:
 $E_{1}(z,\tau) = \zeta(z,\tau) + 2 \zeta(\frac{1}{2})z$.
 $E_{1}(z,\tau)$  is expressed in
 terms of $\theta$-function:
 $E_{1}(z,\tau) = \partial_{z} \ln (\theta(z,\tau))$.

In the elliptic case the matrix of the Lax operator of $r$ point
fusion has the following form:

diagonal part - \be

 L^{ii} = v^{i} +
 (p^{1})^{ii} \cdot \tilde{k}_{2} \cdot E_{1}^{'}(z-x_{a}) + ... +
(p^{r})^{ii} \cdot \tilde{k}_{r} \cdot E_{1}^{(r)}(z-x_{a}) \ee

nondiagonal part - \be

L^{ij} =

(p^{0})^{ij} \cdot \tilde{k}_{1} \cdot \varphi (u^{ij}, z - x_{a})
+ ...

+ (p^{r})^{ij} \cdot \tilde{k}_{r} \cdot  \varphi^{(r)}( u^{ij}, z
- x_{a}),

\ee

where $\tilde{k}_{r}$ is similar to $k_{r}$, commutation relations
are the same as in the rational case. The Hamiltonians expressed
by the variables invariant with respect to the gauge fixing are
found in chapter 3, formula (50), and commutation relations -
formula (53), (54).

\ \ The symplectic structure in the two point fusion case
(rational and elliptic one) is described in chapter 2

The procedure called Inozemtsev Limit ($[8]$) is the method
permits to obtain Toda-like system ( the system with exponential
type of the interaction ) from the elliptic Gaudin system ( the
system with elliptic type of the interaction ) by the combination
of the trigonometric limit, the infinite shift of the particle
coordinates and the renormalization of the interaction constant.
The technical details are given in $[8],[9]$. The Hamiltonians of
the Toda-like system , obtained by Inozemtsev limit from the
elliptic Gaudin system in the case of the two point fusion and
commutation relations were found in chapter 4, formulas (86),(84).
Note, that the other Toda-like systems having nontrivial
commutation relations between the phase space variables were
studied in the work $[10]$.

\section{Systems arising from the rational Gaudin system}

Let us begin from the rational Gaudin system. We will consider the
two point fusion of $n$ marked points ($n > 4$) and fulfil the
following coordinate transformation and decomposition of the
variables $p$: \be

x_{b} = x_{a} + \varepsilon\\
p_{a} = \alpha_{0} p^{0} + \alpha_{1} p^{1} \varepsilon^{-1}, \ \
\ \ p_{b} = \beta_{0} p^{0} + \beta_{1} p^{1} \varepsilon^{-1},

\ee where $\alpha_{r}, \beta_{r}$ are some parameters. Putting
$(16)$ in the matrix of Lax operator $(6)$ and taking the limit
$\varepsilon \rightarrow 0$, we get: \be

L = - \frac{\alpha_{0} p^{0} + \alpha_{1} p^{1}
\varepsilon^{-1}}{(z - x_{a})} + \frac{\beta_{0} p^{0} + \beta_{1}
p^{1} \varepsilon^{-1}}{(z-x_{a}-\varepsilon)} + \sum_{c \neq a,b
} \frac{p_{c}}{(z-x_{c})}.

\ee During the calculation we need to put some additional
conditions on parameters. The reason for their appearance is the
requirement of the absence of singularities coming from degrees of
$\varepsilon^{-1}$ if $\varepsilon \rightarrow 0$. In the present
case we have one condition $\alpha_{1} + \beta_{1} = 0$ and,
setting $\alpha_{0} = 1, \ \ \alpha_{1} = -1, \ \ \beta_{0} = 0, \
\ \beta_{1} = 1$, we get: \be

L_{new} = \frac{p^{1}}{(z - x_{a})^{2}} + \frac{p^{0}}{(z-x_{a})}
+ \sum_{c \neq a,b} \frac{p_{c}}{(z-x_{c})}

\ee The decomposition $<L_{new}^{2}>$ with respect to $(z-x_{a})^{-m}$ has the
following form : \be

<L_{new}^{2}> = H^{2,4}_{1} \cdot (z-x_{a})^{-4} + H^{2,3}_{1}
\cdot (z-x_{a})^{-3} + H^{2,2}_{1} \cdot (z-x_{a})^{-2} +
H^{2,1}_{1} \cdot (z-x_{a})^{-1}\\
+ \sum_{c \neq a,b} ( H^{2,2}_{c} \cdot (z-x_{c})^{-2} +
H^{2,1}_{c} \cdot (z-x_{c})^{-1} )

\ee where: \be

H^{2,4}_{1} =  <(p^{1})^{2}>, \ \ \

H^{2,3}_{1} =  2<p^{1} p^{0}>,\\

H^{2,2}_{1} = <(p^{0})^{2}> + 2\sum_{c \neq a,b} \frac{<p^{1}
p_{c}>}{(x_{a} - x_{c})},\\

H^{2,1}_{1} = \sum_{c \neq a,b} \left[ \frac{<p^{0} p_{c}>}{(x_{a}
- x_{c})} - \frac{<p^{1}p_{c}>}{(x_{a} - x_{c})^{2}} \right], \ \
\

H^{2,2}_{c} = <(p_{c})^{2}> ,\\

H^{2,1}_{c} =  2\frac{<p^{1} p_{c}>}{(x_{c} - x_{a})^{2}} +
\frac{<p^{0} p_{c}>}{(x_{c} - x_{a})} + \sum_{d \neq c,a,b}
\frac{<p_{c} p_{d}>}{(x_{c} - x_{d})},

\ee  where $H^{2,4}_{1}, H^{2,3}_{1}, H^{2,2}_{c}$ are Casimir
operators.

The condition $\sum\limits_{a} p_{a} = 0$, which is the moment
constraint corresponding to the action of the remnant gauge group,
has to change because after the transformation of $p_{a}$ and
$p_{b}$, we get: \be

  p^{0} + \sum_{c \neq a,b} p_{c} = 0,

\ee The numbers of Hamiltonians and Casimir operators do not
change: instead of two Hamiltonians and Casimir operators of the
rational Gaudin system related to the points $x_{a}, x_{b}$, there
appeared $H^{2,4}_{1}, H^{2,3}_{1}$ and $H^{2,2}_{1}, H^{2,1}_{1}$,
 being Hamiltonians and Casimir operators correspondingly. It is
possible to calculate the following commutation relations between the new
variables: \be

\{ (p^{1})^{ij}, (p^{1})^{kl} \} = 0,\\

\{ (p^{0})^{ij}, (p^{0})^{kl} \} = \delta^{il} (p^{0})^{kj} - \delta^{kj} (p^{0})^{il},\\

\{ (p^{0})^{ij}, (p^{1})^{kl} \} = \delta^{il} (p^{1})^{kj} -
\delta^{kj} (p^{1})^{il},

\ee It may be schematically presented as: \be

(p^{1}, p^{1}) \rightarrow 0, \ \ (p^{0}, p^{0}) \rightarrow
 p^{0}, \ \
(p^{0}, p^{1}) \rightarrow
 p^{1},

\ee
  Using the elements $p^{0}, p^{1}$, it is possible to compose
  matrixes of the form (in the case $r=2$): \be

P = \left[
\begin{array}{c}
p^{0}\ \ \ \ \ p^{1}
\\
0\ \ \ \ \ p^{0}
\end{array}
\right]

\ee
 , which generate the parabolic algebra. Taking into account the
position of $p$ in the matrix $P$ (in the case when $p^{r} \in
sl(2,C)$), it is possible to show that the
commutation relations between $p^{r}$ have the following form (Appendix B.): \be

\{ (p^{0})^{ij}, (p^{1})^{kl} \} = \delta^{il} (p^{1})^{k,j+N} -
\delta^{kj} (p^{1})^{i,l+N},\\
 \{ (p^{0})^{ij}, (p^{0})^{kl} \} =
\frac{1}{2} [\delta^{il} ((p^{0})^{kj} + (p^{0})^{k+N,j+N}) -
\delta^{kj} ((p^{0})^{il} + (p^{0})^{i+N,l+N})],

\ee

where $N=2$. The loop decomposition of
 the matrix $P$ in the case of two point fusion gives
 polynomials of the form of $p^{0} + p^{1}\varepsilon^{-1} $. The
 isomorphism can be established by comparing commutation
 relations for the matrix elements and for polynomials.
 Considering the transformation (16) for $x$ and $p$ in the case
 of fusion
 of the first $r$ (for definiteness) of $n$ marked points,
 we get the generalized expression for $L_{new}$, not depending
 on the sequence of those points of fusion but depending (up to the
 coefficients) on the number of points: \be

L_{new} = \frac{p^{r} \cdot k_{r} }{(z - x_{a})^{r}} + ... +
\frac{p^{1} \cdot k_{1} }{(z - x_{a})^{2}} + \frac{p^{0} \cdot
k_{0} }{(z - x_{a})} + \sum_{c \neq a,b} \frac{p_{c}}{(z-x_{c})}

\ee The matrixes $P$, which generate the parabolic algebra, have
the following form: \be

P = \left[
\begin{array}{c}
 p^{0} \ \ \ \ \ p^{1} \ \ \ \ \ ... \ \ \ \ \ p^{n-2} \ \ \ \ \ p^{n-1} \ \ \ \ \ p^{n}
\\
 0 \ \ \ \ \ p^{0} \ \ \ \ \ p^{1} \ \ \ \ \ ... \ \ \ \ \ p^{n-2} \ \ \ \ \ p^{n-1}
\\
 0 \ \ \ \ \ 0 \ \ \ \ \ p^{0} \ \ \ \ \ p^{1} \ \ \ \ \ ... \ \ \ \ \ p^{n-2}
\\
 ... \ \ \ \ \ ... \ \ \ \ \ ... \ \ \ \ \ ... \ \ \ \ \ ... \ \ \ \ \ ...
\\
 0 \ \ \ \ \ 0 \ \ \ \ \ 0 \ \ \ \ \ ... \ \ \ \ \ ... \ \ \ \ \ p^{0}
\end{array}
\right].

\ee Schematically, it is possible to present the commutation
relations in the form of the table: \be

\begin{tabular}{|c|c|c|c|c|c|c|}
\hline
$p$ &  $p^{0}$ & $p^{1}$ & $p^{2}$ & $...$ & $p^{n-1}$ & $p^{n}$\\
\hline
$p^{0}$ & $p^{0}$ & $p^{1}$ & $p^{2}$ & $...$ & $p^{n-1}$ & $p^{n}$\\
\hline
$p^{1}$ & $p^{1}$ & $p^{2}$ & $p^{3}$ & $...$ & $p^{n}$ & $0$\\
\hline
$p^{2}$ & $p^{2}$ & $p^{3}$ & $p^{4}$ & $...$ & $0$ & $0$\\
\hline
$...$ & $...$ & $...$ & $...$ & $...$ & $...$ & $...$\\
\hline
$p^{n-1}$ & $p^{n-1}$ & $p^{n}$ & $0$ & $...$ & $0$ & $0$\\
\hline
$p^{n}$ & $p^{n}$ & $0$ & $0$ & $...$ & $0$ & $0$\\
\hline
\end{tabular}

\ee The intersection of the column and the row is the result of
the commutation.

\paragraph{Symplectic form in the case of the point fusion}

Following works $[2]-[4]$ let us consider the symplectic form of
the Hitchin system in the case of $n$ marked points on $CP^{1}$,
selecting two points for the next fusion: \be

\Omega = \int_{\Sigma} \langle \delta A \wedge \delta \bar{A}
\rangle + \sum\limits_{a} \langle p_{a},g^{-1}_{a}\delta
g_{a} \rangle =\\
=  \int_{\Sigma} [\langle \delta A \wedge \delta\bar{A} \rangle +
\delta(z - x_{a})\langle p_{a},g^{-1}_{a}\delta g_{a} \rangle +
\delta(z - x_{b})\langle p_{b},g^{-1}_{b}\delta g_{b} \rangle +
\sum\limits_{c \neq a,b} \delta(z - x_{c})\langle
p_{c},g^{-1}_{c}\delta g_{c} \rangle],

\ee  where $<>$ is the Killing form, $\delta$ is the exterior
differentiation operator,\\
-$A$ and $\bar{A}$ are holomorphic and antiholomorphic parts of
the
connection correspondingly,\\
-$(p_{a},g_{a}) \in T^{\ast}G_{a}$ where $T^{\ast}G_{a}$ is the
cotangent bundles at the marked point, $p_{a} \in
Lie^{\ast}(G_{a}), g_{a} \in G_{a}$. Symplectic form is invariant
with respect to the action of the group $G_{1}$: \be G_{1} =
\{f(z,\bar{z}) \in C^{\infty}(\Sigma), G \} \ee

The corresponding gauge transformations have the following form:
\be
A = f L f^{-1} +f \partial f^{-1},\\
\bar{A} = f \bar{L} f^{-1} + f\bar{\partial}f^{-1}\\
p_{a} = f_{a} p_{a} f^{-1}_{a}, \ \ g_{a} = g_{a} f^{-1}_{a}. \ee

 The moment map equation
takes the following form: \be

\bar{\partial} L = \delta(z - x_{a})p_{a} + \delta(z - x_{b})p_{b}
+ \sum_{c \neq a,b} \delta(z-x_{c})p_{c}

\ee After transformation (16), we get: \be

\bar{\partial} L = \delta(z - x_{a})p^{0} + \delta'(z -
x_{a})p^{1} + \sum_{c \neq a,b} \delta(z-x_{c})p_{c}.

\ee Solving this equation, we get the Lax operator with the second
order pole. Then let us find such transformation of the form
$\Omega$ which gives us the new form $\Omega_{new}$ corresponding
to (33). Let us introduce the logarithmic coordinates $\ln g$ on
the group $Sl(N,C)$, which are the coordinates on the algebra
 $sl(N,C)$, then $\ln g_{a}
\equiv X_{a}, \ \ \ln g_{b} \equiv X_{b}$ and $g^{-1}_{a}\delta
g_{a} \equiv \delta X_{a}, \ \ g^{-1}_{b}\delta g_{b} \equiv
\delta X_{b}$, and consider the following transformations: \be

X_{a} = X^{0}\varepsilon^{-1} - X^{1}, \ \ \ X_{b} = -
X^{0}\varepsilon^{-1} - X^{1},

\ee Let us define the coupling between the algebra $X$ and the
coalgebra $P$ as $Res_{-1}Tr(PX)$. Substituting (16), putting
$\alpha_{0} = 1, \ \ \alpha_{1} = -1, \ \ \beta_{0} = 0, \ \
\beta_{1} = 1$, and (34) to the form $\Omega$, after expansion in
$\varepsilon$, collecting similar terms and returning to
$X^{0}=(g^{0})^{-1}\delta g^{0}$, we get the new form
$\Omega_{new}$: \be

\Omega = \int_{\Sigma} [\langle \delta A \wedge \delta\bar{A}
\rangle + \delta(z - x_{a})\langle p^{0},g^{-1}_{0}\delta g_{0}
\rangle + \delta^{'}(z - x_{a})\langle p^{1},g^{-1}_{0}\delta
g_{0} \rangle] + \sum\limits_{c \neq a,b} \langle
p_{c},g^{-1}_{c}\delta g_{c} \rangle

\ee It is possible to get (33) from this form.

\section{Systems arising from the elliptic Gaudin system}

\paragraph{Hamiltonians of Gaudin system in the case of two point fusion}
Let us begin (as in the rational case) from the two point fusion
and consider two marked points $x_{a}$ and $x_{b}$. According to
what was said in the introduction, the integrable system is
defined on the symplectic factor-space $R$ ($[4]$) which has the
following form in the case of two points: \be

R = (u,v) \times ({\mathcal O}_{a} \times {\mathcal O}_{b}//H),

\ee where ${\mathcal O}_{a}$ and ${\mathcal O}_{b}$ are the
coadjoint orbits of the group $SL(N,C)$,\\
 $H$ is the Cartan subgroup,\\
$u,v \in \mathcal H$ belong to Cartan subalgebra.

The moment constraint corresponding to the action of the remnant
gauge group has the following view: \be

\sum_{a} p^{ii} = 0

\ee Let us consider the case $N = 2$ (i.e. $p \in sl(2,C)$) and
get the Hamiltonians by decomposing $<L^{2}>$ into the sum of
Eisenstein functions, where $L$ is the matrix of the Lax operator
in the holomorphic representation, with the diagonal part
 \be L^{ii} = v^{i} + \sum_{a} p_{a}^{ii}
\cdot E_{1}(z-x_{a}) \ee

and non-diagonal part  \be

 L^{ij} = \sum_{a} p_{a}^{ij} \cdot \varphi (u^{ij}, z - x_{a}), \ee

where: \be \varphi (u^{ij}, z - x_{a}) =
 \frac{ \theta(u^{ij} + z - x_{a})\theta^{'}(0)}{ \theta(u^{ij}) \theta(z -
 x_{a})}. \ee

In the case of the two point fusion we get:\be

<L^{2}> =
 (v^{1} + p_{a}^{11} \cdot E_{1}(z-x_{a}) + p_{b}^{11} \cdot E_{1}(z-x_{b}))^{2} +\\
 + (v^{2} + p_{a}^{22} \cdot E_{1}(z-x_{a}) + p_{b}^{22} \cdot E_{1}(z-x_{b}))^{2} +\\

 + p_{a}^{12}p_{a}^{21}(-E_{1}^{'}(z-x_{a}) + E_{1}^{'}(u^{12})) +
p_{b}^{12}p_{b}^{21}(-E_{1}^{'}(z-x_{b}) + E_{1}^{'}(u^{12})) +\\

+ 2p_{a}^{12}p_{b}^{21}
 \cdot \varphi (u^{12}, z - x_{a})\varphi (u^{21}, z - x_{b}) + 2p_{a}^{21}p_{b}^{12}
 \cdot \varphi (u^{21}, z - x_{a})\varphi (u^{12}, z - x_{b}).
\ee Analyzing zeroes and poles in $<L^{2}>$ we get the
decomposition in the following form (Appendix C.): \be

<L^{2}> = \sum_{a} ( E_{1}^{'}(z-x_{a}) \cdot H^{2,2}_{a} +
E_{1}(z-x_{a}) \cdot H^{2,1}_{a} ) + H^{2,0}_{a}

\ee , where: \be

H^{2,2}_{a} = <p_{a}^{2}>,

\ee

\be

H^{2,1}_{a} =

 v^{1}p_{a}^{11} + v^{2}p_{a}^{22} + [ p_{a}^{11} p_{b}^{11} + p_{a}^{22}
 p_{b}^{22} ]
  \cdot E_{1} (x_{a}-x_{b}) + \\

+ 2p_{a}^{12}p_{b}^{21}
 \frac{ \theta(u^{12} +  x_{b} -
x_{a}) \theta^{'}(0) }{ \theta(u^{12}) \theta(x_{a} - x_{b})}

+ 2p_{a}^{21}p_{b}^{12}
 \frac{ \theta(u^{12} +  x_{a} -
x_{b}) \theta^{'}(0) }{ \theta(u^{12}) \theta(x_{b} - x_{a})},

\ee

\be

 H^{2,0} =

(v^{1})^{2} + (v^{2})^{2} - [ p_{a}^{11} p_{b}^{11} + p_{a}^{22}
p_{b}^{22} ] \cdot
(E_{1}^{'}(x_{a}-x_{b}) + E_{1}^{2}(x_{a}-x_{b})) +\\

 + [p_{a}^{12}p_{a}^{21} + p_{b}^{12}p_{b}^{21}] \cdot E_{1}^{'}(u^{12})) +\\

 + 2p_{a}^{12}p_{b}^{21} (E_{1}(u^{12}) - E_{1}(u^{12} + x_{b} - x_{a}))
 \cdot
 \frac{ \theta(u^{12} +  x_{b} - x_{a}) \theta^{'}(0) }
 { \theta(u^{12}) \theta(x_{a} -
x_{b})}\\

+ 2p_{a}^{21}p_{b}^{12} (E_{1}(u^{12}) - E_{1}(u^{12} + x_{a} -
x_{b}))
 \cdot
 \frac{ \theta(u^{12} +  x_{a} - x_{b}) \theta^{'}(0) }{ \theta(u^{12}) \theta(x_{b} -
x_{a})}.

\ee Let us calculate the phase space dimension of the integrable
system. There are eight variables: $u^{12}, v^{1}, p_{a}^{11},
p_{a}^{12}, p_{a}^{21}, p_{b}^{11}, p_{b}^{12}, p_{b}^{21}$ in the
system which is in accordance with two Casimir operators and three
Hamiltonians. Fixing the value of Casimir operators we choose the
orbits $H^{2,2}_{a} = (\lambda_{a})^{2}$, $H^{2,2}_{b} =
(\lambda_{b})^{2}$ and taking into account (37) we get: \be

(p_{a}^{11})^{2} + p_{a}^{12}p_{a}^{21}  = (\lambda_{a})^{2}, \ \
\ \ (p_{b}^{11})^{2} + p_{b}^{12}p_{b}^{21}  = (\lambda_{b})^{2},
\ \ \ \ p_{a}^{11} + p_{b}^{11} = 0

\ee Note, that the remnant gauge group in the general case (not
necessarily in the holomorphic representation) consists of the doubly
periodic functions of $(z, \bar{z})$ on the diagonal because they
do not change the gauge fixing in the moment map equation (see
Symmetries). Taking the Fourier-series expansion of these
functions we get the basis in the space of the remnant gauge
transformation. After that let us consider the gauge fixing being
in accordance with the coadjoint action of the remnant gauge group
on $p$ which are independent of $(z, \bar{z})$ diagonal matrices being the
Cartan subgroup with the corresponding moment constraint (37). So
we must consider functions which have only zero harmonic $c_{0} =
\exp(+\alpha)$ in the expansion and generate one-parameter class.
The invariant with respect to the gauge fixing variables take the
form: \be

p_{a}^{12}p_{a}^{21} \equiv x, \ \ p_{b}^{12}p_{b}^{21} \equiv y,
\ \

p_{a}^{12}p_{b}^{21} \equiv z_{1}, \ \ p_{b}^{12}p_{a}^{21} \equiv
z_{2}, \ \

p_{a}^{11}, \ \ p_{b}^{11}, \ \ u, \ \ v

\ee Finally, we get: \be

(p_{a}^{11})^{2} + x = (\lambda_{a})^{2}, \ \ \ (p_{b}^{11})^{2} +
y = (\lambda_{b})^{2}, \ \ \ p_{a}^{11} + p_{b}^{11} = 0, \ \ \ x
y = z_{1} z_{2}

\ee It is clear that six variables (all except $u, v$) can be
expressed through two ones, for example $z_{1}, z_{2}$. So, we get
four-dimensional phase space in accordance with two Hamiltonians.
Note, that the Hamiltonians, as functions on this phase space, are
invariant with respect to the action of the remnant gauge subgroup
because phase space is the factor-space with respect to the action of
this subgroup.

\paragraph{Hamiltonians of the system obtained by the two point fusion}

Let us realize the point fusion in accordance with the
transformation (16). Note, that commutation relations between
$(p^{0})^{ij}$ and $(p^{1})^{ij}$ do not depend on the elliptic
function and are the same as in the rational case. The additional
issue in the evaluation of these limits is the decomposition of
the elliptic function into series of $\varepsilon$. In the case of
$p \in sl(2,C)$ after calculation, setting $u^{12} \equiv u$ and
$v^{1} = - v^{2} = v$, we get (Appendix D.): \be

\frac{1}{2} <L^{2}_{new}> =
 E_{1}^{'''}(z-x_{a}) \cdot H^{2,4} +

 E_{1}^{''}(z-x_{a}) \cdot H^{2,3} +

 E_{1}^{'}(z-x_{a}) \cdot H^{2,2} +

  H^{2,0},

\ee Where : \be

H^{2,4}  = - \frac{1}{6} \cdot \frac{bc}{a} - \frac{1}{6} \cdot
g^{2} = - \frac{1}{12} <(p^{1})^{2}>, \ \ \

H^{2,3}  = \frac{1}{2} \cdot ( b + c ) =  \frac{1}{2} <(p^{0}p^{1})>,\\

 H^{2,2}  = - a + ( c - b ) \cdot E_{1}(u) + \frac{bc}{a} \cdot
[E_{1}^{'}(u) + E_{1}(u)^{2}] - 2vg,\\

H^{2,0}  =

v^{2} +  g^{2} \cdot [\frac{1}{6} E_{1}^{'''}(u) +
(E_{1}^{'}(u))^{2}] + a \cdot E_{1}^{'}(u)

+ ( b - c ) \cdot [ \frac{1}{2} E_{1}^{''}(u) + E_{1}^{'}(u) \cdot
E_{1}(u)] -\\

- \frac{bc}{a} \cdot [\frac{1}{3} E_{1}^{'''}(u) +
(E_{1}^{'}(u))^{2}

+ E_{1}^{''}(u) \cdot E_{1}(u)  +

E_{1}^{'}(u) \cdot E_{1}(u)^{2}],

\ee where we use the following notation for the variables
invariant with respect to the gauge fixing: \be

(p^{0})^{12}(p^{0})^{21} = a, \ \

(p^{0})^{12}(p^{1})^{21} = b, \ \

(p^{1})^{12}(p^{0})^{21} = c, \ \

(p^{1})^{12}(p^{1})^{21} = d = \frac{bc}{a}, \ \

(p^{1})^{11} = g

\ee Let us calculate the dimension of the phase space of the new
system. There are eight variables in the space $(u,v) \times
(p^{r})^{ij}$. $H^{2,4}$ and $H^{2,3}$ are the Casimir operators.
The moment map for the remnant gauge action takes the form
$(p^{0})^{11} = 0$. So, we get the four-dimensional phase space
and two Hamiltonians. Fixing the Casimir operators gives the
equations: \be

 \frac{bc}{a} + g^{2} = \lambda_{0}, \ \ \ \

 b + c = \lambda_{1}

\ee So we get four dimensional phase space and two Hamiltonians.
The Poisson brackets in the new variables take the following form:
\be

\{v,u\} = 1, \ \ \

\{c,b\} = 0, \ \ \

\{a,g\} = c-b,\\

\{c,a\} = -2ag, \ \ \ \{b,a\} = 2ag, \ \ \ \{b,g\} = -
\frac{bc}{a}, \ \ \ \{c,g\} = \frac{bc}{a}. \ee

The Poisson brackets for $d \equiv \frac{bc}{a}$ take the
following form: \be

\{\frac{bc}{a},a\} = 2g(b-c), \ \ \

\{\frac{bc}{a},g\} = 0, \ \ \

\{\frac{bc}{a},b\} = -2g\frac{bc}{a}, \ \ \

\{\frac{bc}{a},c\} = 2g\frac{bc}{a}

\ee

\paragraph{Symmetries}
Hamiltonians defined on the phase space must be invariant with
respect to the remnant gauge action - the Bernstein-Schvartsman
group, which is the semi-direct product of the Weyl group and the
lattice shifts.

 Let us consider (in accordance with [4]) the gauge
transformations preserving the gauge fixing $\bar{L} =
diag(sl[2,C])$, chosen in the moment map equation: \be

\bar{\partial} L + \frac{2 \pi i}{ \tau - \bar{\tau} } \cdot
[\bar{L}, L] =

2 \pi i  \cdot ( \delta^{2}(x_{a})p_{a} + \delta^{2}(x_{b})p_{b}
),

\ee with boundary conditions \be

L(z+1) = L(z), \ \ \ \ L(z+\tau) = L(z), \ee where: \be

\bar{L} = \left[
\begin{array}{c}

 u^{1} \ \ \ \ \ 0
 \\
 0 \ \ \ \ \ u^{2}

\end{array}
\right].

\ee Then the remnant gauge group in the general case consists of
the doubly periodic Cartan valued functions of $(z, \bar{z})$
because they preserve the gauge fixing in the moment map equation.
Taking the Fourier-series expansion of these functions we get
basis in the space of the remnant gauge transformation which
consist of harmonic having the following form: \be

 f^{i} = \exp \left [ 2 \pi i \cdot ( m^{i}
\frac{ z - \bar{z} }{ \tau - \bar{\tau} } +  n^{i}\frac{
\tau\bar{z} - \bar{\tau}z }{ \tau - \bar{\tau} } ) \right ], \ \ \
m^{i}, n^{i} \in Z

\ee Let us consider the transformation of $\bar{L}$ under the basis
function action: \be

2 \pi i \frac{ 1 }{ \tau - \bar{\tau} }

 \left[
\begin{array}{c}

 u^{1} \ \ \ \ \ 0
 \\
 0 \ \ \ \ \ u^{2}

\end{array}
\right]

\rightarrow

f \cdot

2 \pi i \frac{ 1 }{ \tau - \bar{\tau} }

\left[
\begin{array}{c}

 u^{1} \ \ \ \ \ 0
 \\
 0 \ \ \ \ \ u^{2}

\end{array}
\right] \cdot f^{-1} + f \bar{\partial} f^{-1},

\ee

where: \be

f = \left[
\begin{array}{c}

 f^{1} \ \ \ \ \ 0
 \\
 0 \ \ \ \ \ f^{2}

\end{array}
\right],

\ee From this we get law of the transformation for $u^{i}$: \be

 u^{i} \rightarrow u^{i} + m^{i} - n^{i}\tau

\ee Let us consider the transformation of $u^{12}$ in the following
form $u^{12} \rightarrow u^{12} + 1 = u^{1} - u^{2} + 1$. We may
represent this transformation as the following two: $u^{1}
\rightarrow u^{1} + 1$ and $u^{2} \rightarrow u^{2}$. Then $f$ take
the form ($m^{1} = 1, n^{1} = 0$): \be

f = \left[
\begin{array}{c}

 f^{1} \ \ \ \ \ 0
 \\
 0 \ \ \ \ \ 1

\end{array}
\right], \ \ \

 f^{1} = \exp \left [ 2 \pi i \cdot
\frac{ z - \bar{z} }{ \tau - \bar{\tau} }  \right ], \ee It is
possible to define the action of $f$ on $p_{a}$: \be

p_{a} \rightarrow f_{a} p_{a} f_{a}^{-1} =

\left[
\begin{array}{c}

 p_{a}^{11} \ \ \ \ \ p_{a}^{12} \cdot f_{a}^{1}
 \\
 p_{a}^{21} \cdot (f_{a}^{1})^{-1} \ \ \ \ \ p_{a}^{22}

\end{array}
\right]

\ee We obtain the transformation law for $p_{a}^{ij}$: \be

p_{a}^{ii} \rightarrow  p_{a}^{ii}, \ \ \

p_{a}^{12} \rightarrow  p_{a}^{12} \exp \left [ 2 \pi i \cdot
\frac{ x_{a} - \bar{x_{a}} }{ \tau - \bar{\tau} }  \right ], \ \ \

p_{a}^{21} \rightarrow  p_{a}^{21} \exp \left [ - 2 \pi i \cdot
\frac{ x_{a} - \bar{x_{a}} }{ \tau - \bar{\tau} }  \right ]

\ee It is possible to obtain similar expressions for $u^{1}
\rightarrow u^{1} + \tau$, in the general case we have: \be

u^{ij} \rightarrow u^{ij} + m^{i} - n^{i} \tau \\

p_{a}^{ii} \rightarrow  p_{a}^{ii}, \\

p_{s}^{ij} \rightarrow p_{s}^{ij} \cdot \exp \left [ 2 \pi i \cdot
m^{i}\frac{ x_{a} - \bar{x_{a}} }{ \tau - \bar{\tau} } + 2 \pi i
\cdot n^{i}\frac{ \bar{\tau}x_{a} - \bar{x_{a}}\tau }{ \tau -
\bar{\tau} }\right ], \\

\varphi (u^{ij}, z - x_{a}) \rightarrow \varphi ( u^{ij} , z -
x_{a}) \cdot \exp \left [ + 2 \pi i n^{i}\cdot ( z - x_{a}) \right
].

 \ee

\paragraph{Transformations of the system obtained by point fusion}

Let us do the gauge transformations of $L, \bar{L}, p_{s}^{ij}$ in
the moment map equation using the following function: \be

\tilde{f} = \left[
\begin{array}{c}

 \exp \left [- 2 \pi i u^{1} \cdot
\frac{ z - \bar{z} }{ \tau - \bar{\tau} }  \right ] \ \ \ \ \ 0
 \\
 0 \ \ \ \ \ \exp \left [- 2 \pi i u^{2} \cdot
\frac{ z - \bar{z} }{ \tau - \bar{\tau} }  \right ]

\end{array}
\right],

\ee

So we get the moment map equation in the following form:\be

\bar{\partial} L = 2 \pi i  \cdot ( \delta^{2}(x_{a})p_{a} +
\delta^{2}(x_{b})p_{b} ),

\ee with the boundary conditions for the matrix of the Lax operator:\be

L^{ij}(z+1) = L^{ij}(z), \ \ \ \ L^{ij}(z+\tau) = L^{ij}(z) \cdot
\exp \left [- 2 \pi i u^{ij}
  \right ]
\ee Nondiagonal part of the matrix of the Lax operator take the
holomorphic form:\be

 L^{ij} = p_{a}^{ij} \cdot \frac{ \theta(u^{ij} + z - x_{a})
\theta^{'}(0)} { \theta(u^{ij}) \theta(z - x_{a})} + p_{b}^{ij}
\cdot \frac{ \theta(u^{ij} + z - x_{b}) \theta^{'}(0)} {
\theta(u^{ij}) \theta(z - x_{b})}

\ee It is necessary to express the new variables through old ones
$(p_{a})_{new}^{ij} = p_{a}^{ij} \cdot \tilde{f}^{ij}
(\tilde{f}^{ji})^{-1}$ to find the transformations for the new
variables $(p_{a})_{new}^{ij}$ in accordance to the
transformations $u^{ij} \rightarrow u^{ij} + m^{i} - n^{i}
\tau$. Having found the eventual transformation we get it in
the holomorphic gauge: \be

u^{ij} \rightarrow u^{ij} + m^{i} - n^{i} \tau \\

p_{s}^{ij} \rightarrow p_{s}^{ij} \cdot \exp \left [ - 2 \pi
i\cdot n^{i} x_{s} \right ]

 \ee Now we can consider the transformations of the system
obtained by point fusion. Having done the overscaling
transformations of $p_{a}^{ij}$ in the matrix of the Lax operator
in the holomorphic gauge and at the same time the transformations of
the variables under the shift $u^{ij}$, we get the modified action
of the Bernstein-Schvartsman group on the new variables
$(p^{r})^{ij}$, $\varphi (u^{ij}, z- x_{a})$ and its derivative,
having the following form:\be

u^{ij} \equiv u \rightarrow u + m - n\tau,\\

(p^{0})^{^{ij}} \rightarrow (p^{0})^{^{ij}} -

2 \pi i n \cdot (p^{1})^{^{ij}}, \ \ \ \

(p^{1})^{^{ij}} \rightarrow (p^{1})^{^{ij}}, \ \ \ \

(p^{1})^{ii} \rightarrow (p^{1})^{ii},\\

\varphi (u, z - x_{a}) \rightarrow \varphi ( u, z -
x_{a}) \cdot \exp \left [ + 2 \pi i n \cdot ( z - x_{a}) \right ],\\

\varphi^{'} ( u, z - x_{a}) \rightarrow [-2 \pi i n\cdot \varphi (
u , z - x_{a} ) + \varphi^{'} ( u , z - x_{a})] \cdot \exp \left [
+ 2 \pi i n \cdot ( z - x_{a}) \right ]

\ee

\paragraph{Matrix of the Lax operator of the new system}

Let us consider the matrix of the Lax operator in the holomorphic
gauge in the case of two points . Having done transformations
(16) and the decomposition in $\varepsilon$ of the resulting
expression, passing to the limit $\varepsilon \rightarrow
0$ we get:

diagonal part - \be

 L^{ii} = v^{i} - (p^{1})^{ii} \cdot E_{1}^{'}(z-x_{a})

\ee  nondiagonal part - \be

L^{ij} =

(p^{0})^{ij} \cdot \varphi (u^{ij}, z - x_{a})

- (p^{1})^{ij} \cdot  \varphi^{'}( u^{ij}, z - x_{a}) \ee
 The matrix of the Lax operator and the corresponding Hamiltonians
depend on the method of the decomposition of $p^{r}$. It means that
it is possible to consider decomposition in $\varepsilon$ and
$\bar{\varepsilon}$. In the present work we consider
holomorphic decomposition $p^{r}$ parameterized by $\varepsilon$.

 In the
elliptic case with two marked points the total moduli space of the
system is a fibered space with the base defined by the elliptic
module $\tau$. The fiber of this space is a torus with marked
point. This marked point corresponds to a new system having there
a pole of second order in the matrix of Lax operator. This follows
from the considering the module $x_{b}-x_{a}$ (with fixed $\tau$).
In the other words we can fix, for example $x_{a}$ and then $x_{b}$
will take value in the torus. When $x_{b} \rightarrow
x_{a}$ the new system obtained by the fusion of $x_{b}$ and $x_{a}$
corresponds to the fixed point.

Generalizing to the case of $r$ point fusion we get the matrix of
the Lax operator of $r$ point fusion in the following form:

diagonal part - \be

 L^{ii} = v^{i} +
 (p^{1})^{ii} \cdot \tilde{k}_{2} \cdot E_{1}^{'}(z-x_{a}) + ... +
(p^{r})^{ii} \cdot \tilde{k}_{r} \cdot E_{1}^{(r)}(z-x_{a}) \ee

nondiagonal part - \be

L^{ij} =

(p^{0})^{ij} \cdot \tilde{k}_{1} \cdot \varphi (u^{ij}, z - x_{a})
+ ...

+ (p^{r})^{ij} \cdot \tilde{k}_{r} \cdot  \varphi^{(r)}( u^{ij}, z
- x_{a}),

\ee

where $\tilde{k}_{r}$ is similar to $k_{r}$, commutation relations
are the same as in the rational case. Note, that the symplectic
structure is the same as in the rational case, too. Integration in
the formula (35) is taken over the torus.

\section{Inozemtsev Limit}

In the Inozemtsev Limit  procedure we must satisfy the requirement
of the absence of singularities in the integrals of motion which
are the traces of the Lax operators which coincide with the solution of
the moment map equation:\be

\frac{1}{2}<L^{2}>,\ \frac{1}{3}<L^{3}>,\ ..., \frac{1}{k}<L^{k}>

\ee  Taking the Eisenstein-series expansion we get required
 number of Hamiltonians as the coefficients of the decomposition.
Let us write the first nonvanishing term of the Eisenstein
functions in the limit $\omega_2\rightarrow\infty$, taking into
account $u = \tilde{u} + t\omega_2$, where $t$ is some parameter:
\be
\begin{tabular}{|c|c|c|c|c|}
\hline
function & $t = 0$ & $0 < t < 1$ & $t = 1$ & $1 < t < 2$\\
\hline
$E_{1}(u)$
 & $+\frac{1}{2}\frac{\cosh(\frac{\tilde{u}}{2})}{\sinh(\frac{\tilde{u}}{2})}$
 & $+\frac{1}{2}$
 & $+\frac{1}{2}$
 & $+\frac{1}{2}$\\
\hline
$E_{1}^{'}(u)$
 & $-\frac{1}{4}\frac{1}{\sinh^{2}(\frac{\tilde{u}}{2})}$
 & $-e^{-\tilde{u}-t\omega_2}$
 & $-2\cosh(\tilde{u}) \cdot e^{-\omega_2}$
 & $-e^{+\tilde{u}-(2-t)\omega_2}$\\
\hline
$E_{1}^{''}(u)$
 & $+\frac{1}{4}\frac{\cosh(\frac{\tilde{u}}{2})}{\sinh^{3}(\frac{\tilde{u}}{2})}$
 & $+e^{-\tilde{u}-t\omega_2}$
 & $-2\sinh(\tilde{u}) \cdot e^{-\omega_2}$
 & $-e^{+\tilde{u}-(2-t)\omega_2}$\\
\hline
$E_{1}^{'''}(u)$
 & $-\frac{1}{4}\frac{1}{\sinh^{2}(\frac{\tilde{u}}{2})}-\frac{3}{8}\frac{1}{\sinh^{4}(\frac{\tilde{u}}{2})}$
 & $-e^{-\tilde{u}-t\omega_2}$
 & $-2\cosh(\tilde{u}) \cdot e^{-\omega_2}$
 & $-e^{+\tilde{u}-(2-t)\omega_2}$\\
\hline
$...$ & $...$ & $...$ &  $...$ &  $...$\\
\hline $E_{1}^{(2k)}(u)$
 & $+\frac{1}{4}\frac{\cosh(\frac{\tilde{u}}{2})}{\sinh^{3}(\frac{\tilde{u}}{2})}+...$
 & $+e^{-\tilde{u}-t\omega_2}$
 & $-2\sinh(\tilde{u}) \cdot e^{-\omega_2}$
 & $-e^{+\tilde{u}-(2-t)\omega_2}$\\
\hline $E_{1}^{(2k+1)}(u)$
 & $-\frac{1}{4}\frac{1}{\sinh^{2}(\frac{\tilde{u}}{2})}-...$
 & $-e^{-\tilde{u}-t\omega_2}$
 & $-2\cosh(\tilde{u}) \cdot e^{-\omega_2}$
 & $-e^{+\tilde{u}-(2-t)\omega_2}$\\
\hline
\end{tabular}
\ee

Here we put:
\be

 \tau=\frac{\omega_2}{\omega_1},\ \ \omega_1=-i\pi,\ \ \
Im(\omega_2)=0

\ee
 Degeneration of $E_{1}(u)$ up to the second unvanishing
order has the form: \be
E_1(u)=\frac{1}{2}\sum\limits_{k=-\infty}^{\infty}
\coth(\frac{u}{2}-k\omega_2) \rightarrow \left\{
\begin{array}{l}
 0<t<1:\   \frac{1}{2}+\frac{1}{2} e^{-\tilde{u}-t\omega_2}\\
 t=1:\      \frac{1}{2}-\sinh(\tilde{u}) \cdot e^{-\omega_2}\\
 1<t<2:\    \frac{1}{2}-\frac{1}{2} e^{+\tilde{u}-(2-t)\omega_2}
\end{array}
\right. \ee According to this data we will consider the
degeneration of $H^{2,2}$ and $H^{2,0}$. First of all let us
consider some general remarks. We take into account only the first term
in the decomposition $E_{1}(u)$ because the next terms must vanish in order to
avoid the appearance of singularities. So if $0 < t < 2$ we get $E_{1}(u) = \frac{1}{2}$
after the overscaling. The terms in $H^{2,2}$ containing
$E_{1}^{'}(u)$, containing $(E_{1}^{'}(u))^{2}$ in $H^{2,0}$,
and $2vg$ in $H^{2,0}$ also vanish up to requirement of the
absence of singularities after overscaling $H^{2,4}$. The
variables $a, b, c, g$ and $z$ transform in the following way: \be

\tilde{a} = e^{-\chi_a\omega_2}a,\ \ \ \

\tilde{b} = e^{-\chi_b\omega_2}b,\ \ \ \

\tilde{c} = e^{-\chi_c\omega_2}c,\ \ \ \

\tilde{g} = e^{-\chi_g\omega_2}g,\\

z \rightarrow \tilde{z} + s\omega_2

\ee There are the following possible relations between $t$ and some
parameter $s$: \be

s = t, 2-t

\ee Let us consider the case of preserving all Lie-Poisson
brackets between $a,b,c,g$ after overscaling. The conditions of
the absence of singularities in the brackets after overscaling
have the following form: \be

\chi_b \geq \chi_g,\ \ \ \ \chi_c \geq \chi_g,\ \ \ \

\chi_g \geq \chi_b - \chi_a,\ \ \ \ \chi_g \geq \chi_c - \chi_a

\ee The equality sign corresponds to existance of the
Lie-Poisson bracket limit, so: \be

\chi_b = \chi_c = \chi_g, \ \ \ \

\chi_a =0

\ee And the brackets take the following form: \be

\{v,u\} = 1, \ \ \

\{\tilde{c}, \tilde{b}\} = 0, \ \ \

\{a, \tilde{g} \} = \tilde{c}-\tilde{b},\\

\{\tilde{c}, a \} = - 2a\tilde{g},\ \ \ \{\tilde{b}, a \} = +
2a\tilde{g}, \ \ \ \{\tilde{b}, \tilde{g} \} = -
\frac{\tilde{b}\tilde{c}}{a}, \ \ \ \ \{\tilde{c}, \tilde{g} \} =
\frac{\tilde{b}\tilde{c}}{a}

\ee  Casimir operators take the following form: \be

H^{2,4} = \frac{\tilde{b}\tilde{c}}{a} + \tilde{g}^{2}, \ \ \ \

H^{2,3} = \tilde{b} + \tilde{c},

\ee

and Hamiltonians in accordance to the value $t$ take the following form:

\be H^{2,2}=( \tilde{c} - \tilde{b} ) \cdot \frac{1}{2}
 +
\frac{\tilde{b}\tilde{c}}{a } \cdot \frac{1}{2}, \ \ \ \ 0 < t <
2\\

 H^{2,0}= \left\{
\begin{array}{l}
v^{2} - \tilde{g}^{2} \cdot e^{- \tilde{u}} \frac{1}{6}  +
\frac{\tilde{b}\tilde{c}}{a} \cdot e^{- \tilde{u}} \frac{1}{12}
 , \ \ \ \ 0<t<1\\
v^{2} - \tilde{g}^{2} \cdot \cosh(\tilde{u}) \frac{1}{3} -
 ( \tilde{b} - \tilde{c}
) \cdot e^{ + \tilde{u}} + \frac{\tilde{b}\tilde{c}}{a} \cdot
[\frac{13}{12} e^{+ \tilde{u}} + \frac{1}{12} e^{ -
\tilde{u}}], \ \ \ \ t=1\\
v^{2} - \tilde{g}^{2} \cdot e^{ + \tilde{u}} \frac{1}{6} - (
\tilde{b} - \tilde{c} ) \cdot e^{ + \tilde{u}}  +
\frac{\tilde{b}\tilde{c}}{a} \cdot e^{ + \tilde{u}} \frac{13}{12}
, \ \ \ \ 1<t<2
\end{array}
\right. \ee

\section{Appendix A.}
There are main formulas for elliptic functions in this section
(they are borrowed from $[4]$ and $[9]$).
 First of all we define $\vartheta$-function:\be
\vartheta(z,\tau)=q^{\frac{1}{8}}e^{-\frac{\pi}{4}}(e^{i\pi
z}-e^{-i\pi z})
\prod\limits_{n=1}^{\infty}(1-q^{n})(1-q^{n}e^{2\pi iz}) (1-q^n
e^{-2\pi i z}),\ \ q=e^{2\pi i \tau}

 \ee where $\tau$ is the complex module. and Eisenstien functions: \be

E_1(z,\tau)=\partial_{z}log\vartheta(z,\tau),\ \
E_1(z,\tau)\approx \frac{1}{z}+...\\

E_2(z,\tau)=-\partial_z E_1(z,\tau),\ \ E_2(z,\tau)\approx
\frac{1}{z^2}+...\\

\frac{E_2'(u)}{E_2(u)-E_2(v)}=E_1(u+v)+E_1(u-v)-2E_1(u)

\ee Their relations to the Weirstrass functions have the following
form: \be

\zeta(z,\tau)=E_1(z,\tau)+2\eta_1(\tau)z, \ \ \ \
\wp(z,\tau)=E_2(z,\tau)-2\eta_1(\tau).

\ee where $ \eta_1(\tau)=\zeta(\frac{1}{2})$, and $\zeta(z)$ is Riman
$\zeta$-function. Eisenstien functions have the following representations: \be

E_1(z)=\frac{1}{2}\sum\limits_{k=-\infty}^
{\infty}\coth(\frac{z}{2}-k\omega_2), \ \ \
E_2(z)=\frac{1}{4}\sum\limits_{k=-\infty}^{\infty}
\frac{1}{\sinh^2(\frac{z}{2}-k\omega_2)} \ee \be
E_2'(z)=-\frac{1}{4}\sum\limits_{k=-\infty}^{\infty}
\frac{\cosh(\frac{z}{2}-k\omega_2)}{\sinh^3(\frac{z}{2}-k\omega_2)}

\ee There is the following expression in the matrix of the Lax
operator: \be
\varphi(u,z)=\frac{\vartheta(u+z)\vartheta'(0)}{\vartheta(u)\vartheta(z)},

\ee It has a pole at the point $z=0$ and $res|_{z=0}\varphi(u,z)=1$.

The important identity connected with $\varphi$:
\be \varphi(u,v)\varphi(-u,v)=E_2(v)-E_2(u), \ \ \ \
\varphi'_u(u,v)=\varphi(u,v)(E_1(u+v)-E_1(u)) \ee

Behavior on the lattice:
\be

\theta(z+1) = - \theta(z), \ \ \ \ \theta(z+\tau) = -
q^{-\frac{1}{2}}e^{-2 \pi i z}\theta(z),\\
E_1(z+1) = E_1(z), \ \ \ \ E_1(z+\tau) = E_1(z) - 2 \pi i,\\
E_2(z+1) = E_2(z), \ \ \ \ E_2(z+\tau) = E_2(z),\\
\varphi(u+1,z) = \varphi(u,z), \ \ \ \
\varphi(u+\tau,z) = e^{-2 \pi i z}\varphi(u,z).

\ee

\section{Appendix B.}
Let us consider the canonical symplectic form $\omega$ on $T^{*}G
\cong {\mathcal G}^{*} \times G$ ( $G \in SL(N,C)$ ): \be \omega =
\delta <p,g^{-1} \delta g> \ee where $<>$ denotes the Killing form
on Lie(G), $p \in sL^{*}(N,C)$ è $g \in SL(N,C)$, $\delta$ -
external differential operator. We will find the commutation
relation between $F(p,g)$ and $H(p,g)$ defined on $T^{*}G$. There
is Lie-Poisson bracket on $T^{*}G$ ($[11]$):\be

\{F,H\} = C^{jk}_{i}x^{i}\partial_{j}F\partial_{k}H ,

\ee where $C^{jk}_{i}$ is the structure constants of the Lie algebra
 ${\mathcal G}$\\
$\partial_{j} = \frac{\partial}{\partial x^{j}}$ $x^{i}$ is the
coordinate in the space ${\mathcal G}^{*}$.

We want to rewrite (96) in more convenient form for the next
calculations taking into account the explicit dependence of $F$ and
$H$ of $p$ and $g$. Let us represent (96) in the following form:
\be

\{F,H\} = X_{F}H,

 \ee where $X_{F}$ is the vector field corresponding to $F$, so: \be

\{F,H\} = <p, [\frac{\partial F}{\partial p},\frac{\partial
H}{\partial p}]>
 - <g, \{F,H\}_{³}>,

\ee where $\{F,H\}_{³} = \frac{\partial F}{\partial
g}\frac{\partial H}{\partial p} - \frac{\partial F}{\partial
p}\frac{\partial H}{\partial g}$. Put $H = <p \cdot E_{ji}>$ and $F
= <p \cdot E_{lk}>$, so we get the formula (1): \be

 \{F,H\} = p_{kj}\delta^{il} - p_{il}\delta^{kj}

  \ee  In the case of $p \in sl(2,C)$ and two point fusion: \be

G = \left[
\begin{array}{c}
p_{1}^{0}\ \ \ \ \ p_{1}^{1}
\\
0\ \ \ \ \ p_{1}^{0}
\end{array}
\right].

\ee

 Using (98), we get \be

\{ p_{1}^{0,ij}, p_{1}^{1,kl} \} =  \delta^{il}p_{1}^{k,j+N} -
\delta^{kj} p_{1}^{i,l+N},\\
 \{ p_{1}^{0,ij}, p_{1}^{0,kl} \} =
\frac{1}{2} (\delta^{il} p_{1}^{kj} + \delta^{il} p_{1}^{k+N,j+N}
- \delta^{kj} p_{1}^{il} - \delta^{kj} p_{1}^{i+N,l+N}),
 \ee

\section{Appendix C.}
Decomposition of $<L^{2}>$. Put $L$ in the following form: \be

L = \left[
\begin{array}{c}

 L^{11} \ \ \ \ \ L^{12}
 \\
 L^{21} \ \ \ \ \ L^{22}

\end{array}
\right],

\ee then: \be

 \frac{1}{2} TrL^{2} = (L^{11})^{2} + L^{12}L^{21}

\ee Let us consider the decomposition of the diagonal and nondiagonal
parts of $<L^{2}>$ into the Eisenstien functions series in the
case of two point.

\paragraph{Diagonal part}

In this part we will use the property of elliptic functions
with equal periods to differ in constant term if they have equal poles
with the main parts in the parallelogram of periods.
\be

L_{ii} = \sum_{i}[v_{i} + (p_{a})_{ii} \cdot E_{1}(z-x_{a}) +
(p_{b})_{ii} \cdot E_{1}(z-x_{b})]

\ee Let us write the supposed decomposition for $TrL^{2}_{ii}$ (we
do not write $i$, supposing summation): \be

 v^{2} +
 p_{a}^{2} \cdot E_{1}^{2}(z-x_{a}) +
 p_{b}^{2} \cdot E_{1}^{2}(z-x_{b}) +
 2vp_{a} \cdot E_{1}(z-x_{a}) +
 2vp_{b} \cdot E_{1}(z-x_{b}) +
 \\
+ 2p_{a}p_{b} \cdot E_{1}(z-x_{a})E_{1}(z-x_{b}) =\\
=
 - p_{a}^{2} \cdot E_{1}^{'}(z-x_{a})
 - p_{b}^{2} \cdot E_{1}^{'}(z-x_{b})
 + \alpha \cdot E_{1}(z-x_{a})
 + \beta \cdot E_{1}(z-x_{b})
 + c_{0} + v^{2},

\ee where: \be

\alpha = 2p_{a}p_{b} \cdot E_{1}(x_{a} - x_{b}) + 2vp_{a}, \ \ \ \
\beta = 2p_{a}p_{b} \cdot E_{1}(x_{b} - x_{a}) + 2vp_{b},

\ee Now we must define $c_{0}$. In order to do this let us consider $z
\rightarrow x_{a}$, representing $z = x_{a} + \varepsilon$, so as
$\varepsilon \rightarrow 0$. Expressing $c_{0}$ from (84),
decomposing all terms in $\varepsilon$ series, collecting the
similar terms and using the condition $(p_{a})_{ii} + (p_{b})_{ii}=
0$, we get: \be

c_{0} =
 + p_{a}p_{b} \cdot [E^{'}_{1}(x_{a}-x_{b}) +
E^{2}_{1}(x_{a}-x_{b})]

\ee Note that in our calculation $c_{0}$ does not depend on the chosen point (on
the way of calculation).

\paragraph{Nondiagonal part}

In this part we will use the property of elliptic function
with equal periods to differ only by constant multiplier if
they have equal zeros and equal multiplicity of the poles in
the parallelogram of periods.

The term of nondiagonal part of $<L^{2}>$ in the case of two
point is determined in formula (41) of chapter 3. These are the
following expressions having $\theta$-functions: \be

\varphi ( u, z - x_{a})\varphi ( - u, z - x_{a}), \ \ \

\varphi ( u, z - x_{a} )\varphi ( - u, z - x_{b}),

\ee where: \be

\varphi (u^{ij}, z - x_{s}) =
 \frac{ \theta(u^{ij} + z - x_{s})\theta^{'}(0)}{ \theta(u^{ij}) \theta(z -
 x_{s})}

\ee and $s = a,b$. Analyzing zeroes and poles of this
expressions we will extract the elliptic parts, so this enables us to
obtain the decomposition in the Eisenstien functions series
depending on $z - x_{s}$.

\paragraph{Decomposition $\varphi ( u, z - x_{a})\varphi ( - u, z - x_{a})$}

\be

\varphi ( u, z - x_{a} )\varphi ( - u, z - x_{a}) =

\frac{ \theta(u + z - x_{a})\theta( - u + z -
x_{a})(\theta^{'}(0))^{2}}{(\theta(u))^{2}(\theta(z - x_{a}))^{2}}

\ee Zeros: \be

u \rightarrow \pm (z - x_{a})

\ee  Poles: \be

u \rightarrow 0, \ \ \  \varphi ( u, z - x_{a} )\varphi ( - u, z -
x_{a}) \sim -\frac{1}{u^{2}}\\

z \rightarrow x_{a}, \ \ \ \varphi ( u, z - x_{a} )\varphi ( - u,
z - x_{a}) \sim + \frac{1}{ ( z - x_{a} )^{2}}

\ee It follows from the decomposition of $\theta$-functions with respect to small
parameter. Thi decomposition takes the following form: \be

\varphi ( u, z - x_{a} )\varphi ( - u, z - x_{a}) = E^{'}_{1}(u) -
E^{'}_{1}( z - x_{a} )

\ee The similar expression we get for $z - x_{b}$: \be

\varphi ( u, z - x_{b})\varphi ( - u, z - x_{b}) = E^{'}_{1}(u) -
E^{'}_{1}( z - x_{b} )

\ee

\paragraph{Decomposition $\varphi ( u, z - x_{a})\varphi ( - u, z - x_{b})$}

\be

\varphi ( u, z - x_{a} )\varphi ( - u, z - x_{b}) =

\frac{ \theta(u + z - x_{a})\theta( - u + z -
x_{b})(\theta^{'}(0))^{2}}{(\theta(u))^{2}\theta(z -
x_{a})\theta(z - x_{b})}

\ee Zeros: \be

u \rightarrow - (z - x_{a}), \ \ \ u \rightarrow + (z - x_{b})

\ee Poles: \be

u \rightarrow 0, \ \ \ \

\varphi ( u, z - x_{a} )\varphi ( - u, z - x_{b}) \sim
-\frac{1}{u^{2}}\\

z \rightarrow x_{a}, \ \ \ \

\varphi ( u, z - x_{a} )\varphi ( - u, z - x_{b}) \sim + C \cdot
\frac{1}{ ( z - x_{a} )}\\

z \rightarrow x_{b}, \ \ \ \

\varphi ( u, z - x_{a} )\varphi ( - u, z - x_{b}) \sim - C \cdot
\frac{1}{ ( z - x_{b} )}

\ee Using these zeros and poles we may define the combination of
Eisenstien functions: \be

\varphi ( u, z - x_{a})\varphi ( - u, z - x_{b}) = C \cdot [-

E_{1}(z - x_{a}) + E_{1}( z - x_{b} ) - E_{1}(u) + E_{1}(u - x_{a}
+ x_{b} )]

\ee We may define the coefficient $C$ comparing
coefficients in the left and right parts, for example at $z -
x_{b}$: \be

C = \frac{ \theta(u + x_{b} - x_{a})\theta^{'}(0)}{ \theta(u)
\theta(x_{a} - x_{b})}

\ee Finally we get: \be

\varphi ( u, z - x_{a})\varphi ( - u, z - x_{b}) =\\

= \frac{ \theta(u + x_{b} - x_{a})\theta^{'}(0)}{ \theta(u)
\theta(x_{a} - x_{b})}

\cdot [-E_{1}(z - x_{a}) + E_{1}( z - x_{b} ) - E_{1}(u) + E_{1}(u
- x_{a} + x_{b} )]

\ee and for $\varphi ( u, z - x_{b})\varphi ( - u, z - x_{a})$:
\be

\varphi ( u, z - x_{b})\varphi ( - u, z - x_{a}) =\\

= \frac{ \theta(u + x_{a} - x_{b})\theta^{'}(0)}{ \theta(u)
\theta(x_{b} - x_{a})}

\cdot [-E_{1}(z - x_{b}) + E_{1}( z - x_{a} ) - E_{1}(u) + E_{1}(u
- x_{b} + x_{a} )]

\ee

\section{Appendix D.}
Obtaining Hamiltonians of the new system. Let us find: \be

 \frac{1}{2} <L^{2}> = (L^{11})^{2} + L^{12}L^{21},

\ee in the case of the two point fusion. Let us consider
$L^{12}L^{21}$:

\be

 L^{12}L^{21} =\\

 =

[(p^{0})^{12} \cdot \varphi (u, z - x_{a})

- (p^{1})^{12} \cdot  \varphi^{'}( u, z - x_{a})]

 \cdot [(p^{0})^{21} \cdot \varphi ( - u, z - x_{a})

- (p^{1})^{21} \cdot  \varphi^{'}( - u, z - x_{a})]

\ee

We need the next formulas to calculate (123): \be

\varphi ( + u, z - x_{a})\varphi ( - u, z - x_{a}) =

 E_{1}^{'}(u) - E_{1}^{'}(z-x_{a})\\

\varphi^{'}( +u, z - x_{a}) \varphi^{'}( - u, z - x_{a}) =\\

= - [ (E_{1}^{'}(u))^{2} + E_{1}^{2}(u)E_{1}^{'}(u) +
E_{1}(u)E_{1}^{''}(u) + \frac{1}{3}E_{1}^{'''}(u) ] +\\
+ [ E_{1}^{2}(u) + E_{1}^{'}(u) ] \cdot E_{1}^{'}(z - x_{a}) -
\frac{1}{6}E_{1}^{'''}(z - x_{a})\\

\varphi ( + u, z - x_{a}) \varphi^{'} ( - u, z - x_{a}) =\\

= - [ E_{1}(u)E_{1}^{'}(u) + \frac{1}{2}E_{1}^{''}(u) ] +

E_{1}(u) \cdot E_{1}^{'}(z - x_{a}) -

\frac{1}{2}E_{1}^{''}(z - x_{a})\\

\varphi ( - u, z - x_{a}) \varphi^{'} ( + u, z - x_{a}) =\\

= + [ E_{1}(u)E_{1}^{'}(u) + \frac{1}{2}E_{1}^{''}(u) ] -

E_{1}(u) \cdot E_{1}^{'}(z - x_{a}) +

\frac{1}{2}E_{1}^{''}(z - x_{a})

\ee

Put them in (123). Having done the rearrangement and summing with
the decomposition of $(L_{11})^{2} + (L_{22})^{2}$, we get: \be

\frac{1}{2} TrL^{2} =
 E_{1}^{'''}(z-x_{a}) \cdot H^{2,4} +

 E_{1}^{''}(z-x_{a}) \cdot H^{2,3} +

 E_{1}^{'}(z-x_{a}) \cdot H^{2,2} +

  H^{2,0}

\ee Where : \be

H^{2,4} = - \frac{1}{6} (p^{1})^{12}(p^{1})^{21} - \frac{1}{6}
(p^{1})^{11}(p^{1})^{11}, \ \ \ \

 H^{2,3} = \frac{1}{2}[ (p^{0})^{12}(p^{1})^{21} + (p^{0})^{21}(p^{1})^{12} ]

\ee

\be

H^{2,2} =

 - (p^{0})^{12}(p^{0})^{21} +
 E_{1}(u) \cdot
[ - (p^{0})^{12}(p^{1})^{21} + (p^{0})^{21}(p^{1})^{12} ]+\\

+ (p^{1})^{12}(p^{1})^{21} \cdot [ E_{1}^{'}(u) + E_{1}^{2}(u)]

- 2v(p^{1})^{11}

\ee

\be

H^{2,0} = v^{2} + ((p^{1})^{11})^{2} \cdot [\frac{1}{6}
E_{1}^{'''}(u) + (E_{1}^{'}(u))^{2}]

+ (p^{0})^{12}(p^{0})^{21} \cdot E_{1}^{'}(u) +\\

+ [(p^{0})^{12}(p^{1})^{21} - (p^{0})^{21}(p^{1})^{12}] \cdot

[E_{1}(u)E_{1}^{'}(u) + \frac{1}{2}E_{1}^{''}(u)

] -\\

- (p^{1})^{12}(p^{1})^{21} \cdot

[(E_{1}^{'}(u))^{2} + E_{1}^{2}(u)E_{1}^{'}(u) +
E_{1}(u)E_{1}^{''}(u) + \frac{1}{3}E_{1}^{'''}(u)]

 \ee

\paragraph{Acknowledgments}
Author would like to thank A. Chervov, V. Dolotin, A. Mironov, D.
Talalaev and A. Zotov for the useful remarks, A. Levin for
fruitful and numerous discussions. Author would like to especially
thank M. Olshanetsky for the suggested theme, support and
attention during the writing of article. The work was supported in
part by grant RFFI 00-02-16530 and grant 00-15-96557 for support
of scientific schools.


\begin{thebibliography}{0}

\bibitem{1}
Gaudin, J. Phys. (Paris)37, 1087 (1976).

\bibitem{2}
N.Nekrasov. Holomorphic bundles and many-body
systems.hep-th/9503157.

\bibitem{3}
B.Enriques and V.Rubtsov, Math.Res.Lett. 3 (1996), 343-357.

\bibitem{4}
A.M. Levin and M.A. Olshanetsky. Hierarchies of isomonodromic
deformations and Hitchin systems.hep-th/9709207.

\bibitem{5}
N.Hitchin. Duke Math.J.1987.V.54. ü 1.p.91.

\bibitem{6}
A. Beauville. Acta Math. 164 No.3/4 (1990) 211-235

\bibitem{7}
A. Chervov, D. Talalaev.
Hitchin systems on singular curves II. Gluing subschemes. hep-th/0309059.

\bibitem{8}
V.I. Inozemtsev. Commun.Math.Phys.1989. V.121. p.629.

\bibitem{9}
Eric D'Hoker and D.H. Phong. Lectures on supersymmetric Yang-Mills
theory and integrable systems.hep-th/9912271.

\bibitem{10}
Y.Chernyakov, A.Zotov, TMPH, 129 ü 2, 2001, 258-277.

\bibitem{11}
A.M. Perelomov. Integriruemie systemi clasicheskoy mehaniki i
algebri Lie. M., Nauka, 1990.


\end{thebibliography}
\end{document}